\documentclass[useAMS,usenatbib]{mn2e}

\usepackage{epsfig}
\usepackage{longtable}
\usepackage{times}

\usepackage{amsmath}
\usepackage{amssymb}

\def \vela {Vela\,X--1}

\def \inte {\textit{INTEGRAL}}

\def \degmark {^\circ}

\def \hcm {\hbox {\ifmmode $ atom cm$^{-2}\else atom cm$^{-2}$\fi}}

\def \apj {ApJ}
\def \aj {AJ}
\def \apjl {ApJL}
\def \apjs {ApJS}
\def \aap {A\&A}

\def \pasj {PASJ}

\def \mnras {MNRAS}

\newcommand{\be}{\begin{equation}}

\newcommand{\ee}{\end{equation}}

\title[Wind structures in Vela X--1 with INTEGRAL]{Probing large-scale wind structures in Vela X--1 
using off-states with $INTEGRAL$}
\author[Sidoli et al.]{L.~Sidoli,$^{1}$\thanks{E-mail: sidoli@iasf-milano.inaf.it} A.~Paizis,$^{1}$ F. F\"urst,$^{2}$  J.~M.~Torrej\'{o}n,$^{3}$  
 P.~Kretschmar,$^{4}$   E.~Bozzo,$^{5}$ and \newauthor K.~Pottschmidt$,^{6, 7}$ \\
$^{1}$INAF, Istituto di Astrofisica Spaziale e Fisica Cosmica, Via E.\ Bassini 15,   I-20133 Milano,  Italy   \\
$^{2}$Cahill Center for Astronomy and Astrophysics, California Institute of Technology, Pasadena, CA 91125, USA \\
$^{3}$Instituto Universitario de F\'{\i}sica Aplicada a las Ciencias y las Tecnolog\'{\i}as, Universidad de Alicante, E-03690 Alicante, Spain \\
$^{4}$European Space Astronomy Centre (ESA/ESAC), Science Operations Department, E-28691 Villanueva de la Ca\~nada (Madrid), Spain \\ 
$^{5}$INTEGRAL Science Data Centre, Universit\'e de Gen\`eve, Chemin d'\'Ecogia 16, 1290 Versoix, Switzerland \\
$^{6}$Center for Space Science and Technology, University of Maryland Baltimore County, Baltimore, MD 21250, USA \\
$^{7}$CRESST and NASA Goddard Space Flight Center, Astrophysics Science Division, Code 661, Greenbelt, MD 20771, USA \\
}

\begin{document}

\date{Accepted 28 November 2014. Received 28 November 2014; in original form 17 October 2014}

\pagerange{\pageref{firstpage}--\pageref{lastpage}} \pubyear{2014}

\maketitle

\label{firstpage}

\begin{abstract}
Vela X--1 is the prototype of the class of wind-fed accreting pulsars 
in high mass X--ray binaries hosting a supergiant donor. 
We have analysed in a systematic way ten years of \inte~data of \vela\  (22--50\,keV) and we 
found that when outside the X--ray eclipse, the source undergoes several luminosity drops
where the hard X--rays luminosity goes below $\sim$3$\times$10$^{35}$~erg~s$^{-1}$,
becoming undetected by \inte.
These drops in the X--ray flux
are usually referred to as ``off-states'' in the literature.
We have investigated the distribution of these off-states along the \vela\ $\sim$8.9~d orbit,
finding that their orbital occurrence displays an asymmetric distribution, 
with a higher probability to observe an off-state near the pre-eclipse than during the post-eclipse.
This asymmetry can be explained by  scattering of hard X--rays 
in a region of {\it ionized} wind, able to reduce the source hard X--ray brightness
preferentially near eclipse ingress. 
We associate this ionized large-scale wind structure with the photoionization wake 
produced by the interaction of the supergiant wind with the X--ray emission from the neutron star.
We emphasize that this observational result could be obtained thanks to the accumulation 
of a decade of \inte\ data, with observations covering 
the whole orbit several times, allowing us to detect an asymmetric pattern in the 
orbital distribution of off-states in \vela.

\end{abstract}


        \section{Introduction\label{intro}}

\vela\ is an eclipsing and detached high-mass X--ray binary (HMXB) hosting an X--ray pulsar (P$_{spin}$$\sim$283~s, \citealt{McClintock1976}) 
that accretes matter from the wind of its B0.5Ib companion HD 77581 (see \citealt{Rawls2011} and references therein).
Located at a distance of 1.9 kpc \citep{Sadakane1985}, the orbital period of the system is 8.96~d (\citealt{vanKerkwijk1995}, 
\citealt{Kreykenbohm2008}), implying that the neutron star orbit is embedded in the companion wind.

The structure of the supergiant wind has been studied at different wavelengths, 
by means of ultraviolet and optical spectroscopy (\citealt{Kaper1994}, \citealt{vanLoon2001})
and  X--ray spectroscopic studies at different orbital phases 
(\citealt{Eadie1975}, \citealt{Nagase1986}, \citealt{Sato1986}, \citealt{Haberl1989}, \citealt{Lewis1992}, 
\citealt{Sako1999}, \citealt{Schulz2002}, \citealt{Goldstein2004},  \citealt{Watanabe2006}, 
\citealt{Furst2010}, \citealt{Doroshenko2013}, \citealt{Martinez-nunez2014}
and references therein).
These studies indicate the presence of both cold and hot gas components in the system, consistent with 
photoionization of the  stellar wind produced by the X--ray pulsar, leading to the
formation of a so-called photoionization wake trailing the neutron star 
(\citealt{FF1980}, \citealt{Kaper1994}, \citealt{Feldmeier1996}, \citealt{vanLoon2001}), 
confirmed  by simulations (\citealt{Blondin1990}, \citealt{Blondin1991}, \citealt{Mauche2008}). 

The X--ray emission of \vela\ is persistent at a level of 10$^{36}$~erg~s$^{-1}$, 
although variable (within a factor of $\sim$10),
showing rare giant flares together with so-called off-states, that manifest themselves 
as flux drops lasting a few pulse periods 
(\citealt{Kreykenbohm2008}, \citealt{Doroshenko2011}). 
The observed variability is indicative of the presence of massive clumps in the supergiant 
wind (\citealt{Nagase1986}, \citealt{Furst2010}). 
The global variability of \vela\ 
at hard X--rays with \inte\ has been studied by different authors (\citealt{Furst2010}, \citealt{Paizis2014}),
pointing to a log-normal luminosity distribution.

Here, we present the first study of the orbital dependence of the {\em off-states} in \vela, 
as observed in a decade of \inte\ observations (22--50 keV).

\begin{figure*}
\begin{center}
\centerline{\includegraphics[width=16cm,angle=0]{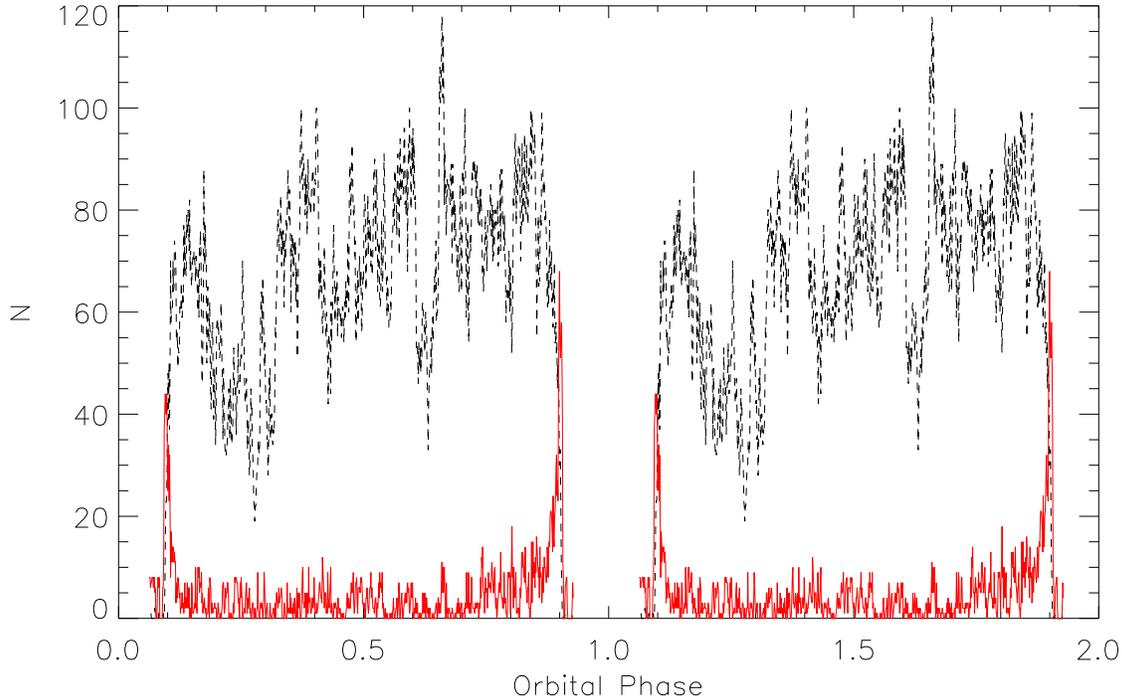}} 
\caption{Histogram of the orbital distribution of the 100~s off-states in Vela X--1 (red solid curve, bin size $\Delta \phi$ = 0.001, that is $\sim$774~s), 
compared to the orbital distribution of the detections (black dashed curve). Data are in the energy band 22--50 keV. Two orbits are shown
for clarity. Eclipses are evident around $\phi$=0, 1, 2, where \vela\ is not detected by \inte\ (see text for details). 
}
\label{fig:off}
\end{center}
\end{figure*}

  	 \section{Data Analysis \label{OSA}}
In orbit for more than a decade, \inte~enables an important long-term study 
of the hard X--ray properties of high energy sources. In this paper, we 
used \inte~archival public data of \vela, from the beginning of the 
mission in 2002, up to revolution 1245, for a total of about 10.2 years of 
data.  We have considered only the pointings in which \vela~was within 
$12\degmark$~from the centre and with a duration $>$1\,ks.
The raw data were downloaded from the ISDC Data Centre for Astrophysics 
into our local \inte/IBIS data base \citep{Paizis2013}. Standard analysis 
has been applied using OSA 10.0 software 
package\footnote{http://www.isdc.unige.ch/integral/analysis}.

We produced individual pointing images ($\sim$ks) and the associated detected source 
lists in the 22--50\,keV energy band, as well as light-curves binned over 
100\,s in the 22--50\,keV band. The energy range was chosen to maximize 
the signal-to-noise ratio while minimizing the instrumental low threshold 
fluctuations \citep{Caballero2012}. 
We consider a threshold of 5$\sigma$ for a detection in the 
imaging analysis and of 3$\sigma$ for a detection in the lightcurve 
analysis. 
In the obtained data set, \vela~is within $12\degmark$~from the centre for about
4.84\,Ms, and it is detected at the imaging level in about 3.98\,Msec
($\sim$82\% of the time).

The analysis of the \vela\ cumulative luminosity distribution of
a significantly overlapping data set (9 years of \inte\ data, instead of 10.2 years) 
has been reported in \citet{Paizis2014}. 
We refer to that paper for more details on the  hard X--ray emission properties 
of Vela X--1, as observed by \inte.
Here, we concentrate on the orbital distribution of the source's off-states.

	      \section{Results}\label{sec:res}

We considered  all \inte\ data (spanning about 10 years) where \vela\ was detected at the 
imaging level ($\sim$ks). 
Indeed, non-detections at imaging level are found only during X-ray eclipses.
We then extracted the hard X--ray light curves (22--50 keV) from these pointings, 
adopting a  bin time of 100~s.

We found that, outside the eclipses, the source was not detected by \inte\ (implying a hard 
X--ray observed luminosity, L$_{X}$$\lesssim$3$\times$10$^{35}$~erg~s$^{-1}$) 
during short time intervals 
(lasting from 100 s to a few hundred seconds). 
We will refer to these non-detections
as  ``off-states'' or ``dips'', regardless of the cause of diminished flux 
(similarly to the definition assumed by \citealt{Kreykenbohm2008}, who studied
a subsample of \inte\ data of \vela, at early times of the mission).

Assuming the ephemeris reported in \citet{Kreykenbohm2008}, we studied the distribution 
of these off-states along the orbit.  
In Fig.~\ref{fig:off} we compare the off-states distribution with 
the occurrences of the source detections (both at 100 s level) along the orbit.
We found two remarkable features in the off-state distribution (red solid curve in Fig.~\ref{fig:off}): 
the first is that off-states  occur at any orbital phase; 
the second and, more important, is that they cluster near the eclipse, 
but with an asymmetric profile: at late orbital phases, approaching the eclipse, 
the off-states are more numerous and span a broader phase interval 
than during the eclipse egress, where the peak 
in the off-states distribution is  narrower.

\begin{figure}
\begin{center}
\centerline{\includegraphics[width=6.1cm,angle=-90]{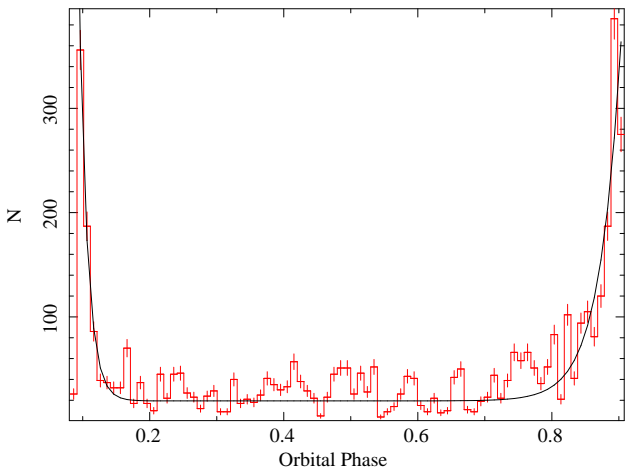}} 
\caption{Histogram of the orbital distribution of the 100~s off-states 
in Vela X-1 (bin size $\Delta \phi$ = 0.01, that is $\sim$7740~s), 
fitted with a constant and two exponential functions. The exponents
of the two exponential functions ($e^{-\alpha_{1}\phi }$ and $e^{+\alpha_{2}\phi}$) 
describing the enhanced frequency of off-states near eclipse egress and  eclipse ingress are 
significantly different (see text).}
\label{fig:fit}
\end{center}
\end{figure}

To quantify this effect, we rebinned the off-states orbital distribution 
adopting a bin size of $\Delta \phi$ = 0.01,
and fitted this distribution (assuming an uncertainty on the $N$~axis of $\sqrt{N}$) 
with a model consisting of two exponential functions 
(proportional to $e^{-\alpha{_1}\phi}$ and $e^{+\alpha_{2}\phi}$) 
together with a constant function, finding two significantly 
different exponents: $\alpha_1$ = $32\pm{2}$   
and  $\alpha_2$= $79\pm{4}$,
near eclipse egress ($\phi$=0.1-0.2) and  eclipse ingress ($\phi$=0.7-0.9), respectively (Fig.~\ref{fig:fit}).

Our analysis demonstrates that two different types of off-states co-exist in \vela: 
the ones uniformly distributed along the orbit 
and the ones more concentrated near the X--ray eclipse, 
with a broader coverage in orbital phases near eclipse ingress than near eclipse egress.

The first kind of dips occurring at any orbital phase 
can be explained by intrinsic X--ray variability that causes the source to be un-detected by \inte.
When observed with more sensitive instruments below 10 keV, off-states usually show a spectral softening,
thus suggesting that they are not simply due to obscuration by a dense wind clump passing in front of
the X--ray pulsar, but by a change in the accretion regime or by the onset of a propeller effect.
These dips  have been  studied in detail by several authors 
(e.g. \citealt{Kreykenbohm2008}, \citealt{Doroshenko2011}, \citealt{Doroshenko2012}, 
\citealt{Shakura2013}, \citealt{Odaka2013}, \citealt{Furst2014}) and 
have been found also in other accreting X--ray pulsars 
(even in Supergiant Fast X--ray Transients, e.g. \citealt{Drave2014}).

The second kind of dips, asymmetric and clustered 
around the eclipses, could be discovered only 
by taking advantage of the long-term observations of Vela~X--1 
available with IBIS/ISGRI. This result
points to a different mechanism, not intrinsic to the X--ray source. 
Indeed, the orbit is almost circular and cannot induce 
any variability in the accretion rate leading to dips
preferentially seen at certain phases. 
Moreover, eclipse in \vela\ occurs near periastron (e.g. Fig.~1 in \citealt{Martinez-nunez2014}). 

More likely, the two peaks in the off-states distribution near $\Delta$$\phi$=0.1 and 0.9
probe the innermost denser regions of the supergiant wind. The larger orbital extent
of dip occurrence in the pre-eclipse region can be induced by the passage into the line of sight 
of a large-scale {\it ionized} wind structure, 
able to produce a drop in the X--ray flux by scattering.
This structure can be naturally associated 
with the photoionized wake (see below; e.g. \citealt{Kaper1994}).

	      \section{Discussion}\label{sec:discussion}

Dips in the light curve of \vela\ have been observed since 
the early 1970s: in two orbital cycles observed with $Ariel~V$ (1.2-19.8 keV) 
four irregular dips were detected
(with two dips at similar late orbital phases in both cycles) 
and an association with the obscuration by an accretion wake was proposed 
(\citealt{Eadie1975}, \citealt{Charles1978}). 

Simulations by \citet{Blondin1990, Blondin1991} have shown that three different wind
structures are present in the stellar wind, excited by the presence of the accreting neutron
star: the {\em accretion wake}, the {\em tidal stream} and the {\em photoionization wake}. 
A sketch of these three wind structures in Vela X--1 can be found in \citet{Kaper1994}.
The {\em accretion wake} around the neutron star can affect the 
column density variations along the orbit only around $\phi$=0.4--0.5 orbital phases.
The {\em tidal stream} is a permanent wind enhancement structure that 
is a source of strong orbital phase-dependent attenuation of soft 
X--rays, 
expected to produce a higher hydrogen column density  
from phase $\phi$=0.5 up to the X–-ray eclipse at phases $\phi$=0.9-1.1. 
The production of the tidal stream depends on the orbital separation, being
more evident in closer orbits \citep{Blondin1991}. 
An enhancement of the neutral absorption column density
towards the neutron star at late orbital phases has been indeed observed in \vela\ 
(e.g. \citealt{Nagase1986}, \citealt{Doroshenko2013} and references therein).
Finally, the wind structure is strongly influenced by the neutron star 
orbit not only because of gravity but also because of the accretion 
X--ray emission produced by the compact object, 
that photoionizes the wind reducing the ability of the wind to be 
radiatively driven \citep{Castor1975}.
This effect decreases the wind velocity near the neutron star and produces a wake of gas
trailing behind the ionized wind, where faster wind collides with slower wind,
leading to the formation of a so-called {\em photoionization wake} (\citealt{FF1980}, \citealt{Blondin1990}, \citealt{Feldmeier1996}).
Evidence for a  photoionization wake in \vela\ has been found also from optical 
spectroscopy \citep{Kaper1994}. 

The orbital distribution of off-states we observed with \inte\ is  consistent  with
the orbital phases spanned by the tidal stream discussed by \citet{Blondin1991} 
and by the photoionization wake \citep{Kaper1994}, that somehow overlap.

Thanks to the accumulation of a huge amount of \inte\ data,
able to cover the entire orbit several times, we could study the 
orbital dependence of the occurrence of the off-states, finding an asymmetric distribution.
We explain it with the presence of a large-scale {\em ionized} wind structure that scatters hard X--rays
out of the line of sight, producing short off-states.

Scattering by the ambient wind is also the physical mechanism explaining both
the residual X--ray emission observed during the eclipse and the soft X--rays excess visible in \vela\ spectrum
(\citealt{Haberl1990, Haberl1991}, \citealt{Lewis1992}, \citealt{Feldmeier1996}). 

Our finding of an asymmetry in the off-state distribution, with an enhanced number 
of off-states at late orbital phases, complements
previous studies on the circumstellar matter in \vela,
starting from \citet{Eadie1975}, \citet{Nagase1986}, \citet{Haberl1989}, \citet{Lewis1992}
to the most recent works on the
neutral hydrogen column density 
variations along the orbit as observed by ASM/$RXTE$ \citep{Furst2010} 
and MAXI data \citep{Doroshenko2013} 
in much softer X--ray bands. 
A  hardening of the X--ray emission 
at late orbital phases was found by these authors, explained   
with denser gas into the line of sight due to the presence of a photoionization wake,
with a column density, N$_{H}$=1-3$\times$10$^{23}$~cm$^{-2}$.
Here, we were able to map the ionized component of this gas stream 
that produces a significant scattering effect at hard X--rays, outside the eclipse,
only detectable after the accumulation of a decade of \inte\ observations.

We note that a similar effect was observed in \inte\ hard X--ray data of Cyg~X--1: 
this HMXB black hole was simultaneously observed at soft and hard X--rays, finding that   
dips are mainly present in the soft X--ray band at upper
conjunction due to photoelectric absorption in the focused stellar wind,
whereas in \inte\ simultaneous data, a scattering effect is evident \citep{Hanke2010}.
These authors suggested that these features are due to clumps in the focused wind which 
are both highly ionized and with (near-)neutral cores. The neutral clump cores produce dips
by photoelectric absorption in the soft X--rays, while their ionized halo produces the attenuation
at hard X--rays. 
We conclude here that a similar scattering effect is indeed at work in \vela\ too.

\begin{figure}
\begin{center}
\centerline{\includegraphics[width=8.5cm,angle=0]{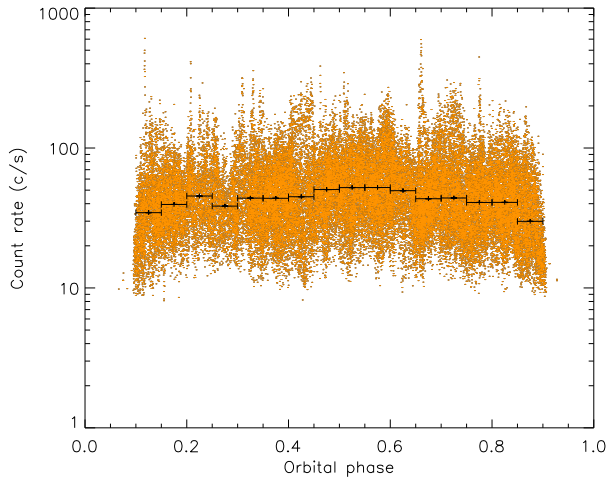}} 
\caption{Long-term orbital light curve of Vela X--1 at 
hard X--rays (22--50 keV, 10.2 years of \inte\ data) outside the eclipses.
The orange points show the Vela X--1 detection count rates in 100~s bins.
Uncertainties on the source count rates are not shown, 
for clarity (average error on count rate is about 4~counts~s$^{-1}$).
The thick black line marks the median count rate in each phase bin.}
\label{fig:good_orbit}
\end{center}
\end{figure}

For completeness, we show in Fig.~\ref{fig:good_orbit} the 
long-term orbital light curve of Vela X--1 at hard X--rays (22--50 keV), 
as obtained by IBIS/ISGRI in this work. 
The curve is consistent with 
what found by \citet{Furst2010} who analysed a more 
contained \inte\ data-set in the similar energy band 20--60 keV. 
The comparison of this long-term folded light curve with the
off-states orbital distribution (Fig.~1) shows that the pattern we discovered and present in this work 
is clearly visible \textit{only plotting the off-states distribution}, rather than 
the more standard  orbital light curve that is dominated by the source
variability.

\section*{Conclusions}

We have analysed about 10 years of \vela\ \inte\ data.
We have reported here on the first evidence at hard X--rays (22--50 keV) of an
orbital dependence of the off-states distribution in \vela\ (time intervals when
the source is undetected by \inte\, i.e. L$_{X}$$\lesssim$3$\times$10$^{35}$~erg~s$^{-1}$).
Their orbital distribution is asymmetric:
the off-states cluster near the X--ray eclipse, with a broader orbital distribution
before the eclipse than after it, covering the orbital phase range $\phi$=0.7-0.9. 

We exclude an intrinsically fainter X--ray flux producing more non-detections
with \inte\ at these orbital phases, given the orbital (almost circular) geometry. 
Moreover, the periastron, and not the apastron, is located near the eclipse phases, thus potentially 
producing the opposite effect of an enhanced X--ray luminosity.

More likely, the orbital asymmetry of the off-states distribution is
indicative of a higher density of ionized material 
trailing the neutron star along its orbit, causing 
an attenuation of hard X--ray flux into the line of sight at late orbital phases. 
We associated this extra  scattering with an ionized large-scale wind structure,
very likely the so-called photoionization wake,  which is expected to lie at similar orbital phases.

Our analysis demonstrates that \inte\ archival observations, mapping the off-states
distribution along the \vela\ orbit many times, can provide meaningful information 
on the structure of the supergiant {\it ionized} wind at large scales.

\section*{Acknowledgements}

%
Based on observations with \textit{INTEGRAL}, an ESA project
with instruments and science data centre funded by ESA member states
(especially the PI countries: Denmark, France, Germany, Italy,
Spain, and Switzerland), Czech Republic and Poland, and with the
participation of Russia and the USA. 
This work has made use of the \inte~archive developed at INAF-IASF Milano, 
http://www.iasf-milano.inaf.it/$\sim$ada/GOLIA.html.
We acknowledge support from ISSI 
through funding for the International Team 
on ``Unified View of Stellar Winds in Massive X-ray Binaries'' (PI: S. Mart\'inez-Nu$\tilde{\rm n}$ez).
LS thanks L.~Oskinova for interesting discussions.
JMT acknowledges grant AYA2010-15431.
LS and AP acknowledge
the Italian Space Agency financial support INTEGRAL
ASI/INAF agreement no. 2013-025.R.0.

\bsp

\label{lastpage}

\end{document}